\documentclass{aa}

\usepackage{graphics,times}

\def\kms{\hbox{km\,s$^{-1}$}}
\def\mag{\hbox{$^{\rm m}$}}

\begin{document}

\thesaurus{06(08.09.2 WRA\,751; 08.05.3; 08.13.2; 09.02.1; 09.10.1)}

\title{A kinematic and morphological investigation of the asymmetric \\ 
nebula around the LBV candidate WRA\,751}

\titlerunning{The asymmetric nebula around WRA\,751}

\author {K.\ Weis
\inst{1,2,}
\thanks{Visiting Astronomer, Cerro Tololo Inter-American Observatory,
National Optical Astronomy Observatories, operated by the Association of
Universities for Research in Astronomy, Inc., under contract with the National
Science Foundation.}
}

\offprints{K.\ Weis, Institut f\"ur Theoretische Astrophysik, Tiergartenstr.
15, 69121 Heidelberg, Germany - E-mail: kweis@ita.uni-heidelberg.de}

\mail{K.\ Weis, Institut f\"ur Theoretische Astrophysik, Tiergartenstr. 15,
69121 Heidelberg, Germany}

\institute{Institut f\"ur Theoretische Astrophysik, Tiergartenstr. 15, 69121
Heidelberg, Germany \and Max-Planck-Institut f\"ur Radioastronomie, Auf dem
H\"ugel 69, 53121 Bonn, Germany }

\date{Received / Accepted}

\maketitle

\begin{abstract}

WRA\,751 is an evolved massive star in our Galaxy closely resembling Luminous
Blue Variable stars (LBVs). It is surrounded by a nitrogen enriched nebula of about
23\arcsec\ diameter. A comparative study of the nebula's morphology and
kinematics is presented, it supports---together with spectroscopical
evidence---the classification of WRA\,751 as a LBV. Images show that the nebula
consists of a nearly spherical shell as well as a bipolar-like structure north
and south of its main body, the Northern and Southern Caps.

In contrast to the almost spherical appearance of the main body of
the nebula, the kinematics shows a deviation even of this part
from a classical spherical expansion pattern. From the present
data it can be concluded that the main body expands asymmetrically
(central expansion velocity $\sim 26$\,\kms), with a thicker shell
at the back side. A bump-like structure can be found to the west
of the central star. In addition to the main body, bipolar
kinematic components can be identified with the morphologically
classified Caps. These results put WRA\,751 into the class of LBVs
which are surrounded by a nebula with bipolar components,
albeit considerably less pronounced than, for instance, in the
classical bipolar LBVs $\eta$ Car and HR Car.


\end{abstract}

\section{Introduction}

With today's initial mass function, stars with masses of $M \sim
50 - 100$\,M$_{\sun}$ and luminosities of $L\sim
10^{5-6}$\,L$_{\sun}$ top the {\it Hertzsprung-Russell diagram\/}
(HRD) and represent the most massive stars known. These massive
stars show a remarkable evolutionary behavior (e.g., Schaller
et al.\ 1992, Stothers \& Chin 1996). After spending a `normal'
life as O stars on the main sequence they evolve towards cooler
temperatures while entering a phase with very high mass loss (up
to 10$^{-4}$M$_{\sun}$\,yr$^{-1}$), they become {\it Luminous Blue
Variables\/} (LBVs). This phase starts when the stars reach the
{\it Humphreys-Davidson limit\/} (Humphreys \& Davidson 1979,
1994, Langer 1994) in the HRD. Analyzing HRDs of the Galaxy and
the LMC, Humphreys (1978, 1979) and Humphreys \& Davidson (1979)
found a lack of very luminous red supergiants. Apperently the most
massive stars do not evolve into red supergiants but instead their
evolution is reversed towards the blue supergiant part in the HRD.
The turning points form the empirical Humphreys-Davidson limit. 
Around this Humphreys-Davidson limit in the HRD LBVs are found. One of the 
most prominent characteristics of the unstable LBV phase are 
strong stellar winds and possible giant eruptions, which lead to the 
peeling off of parts of the stellar envelope and the formation of small
circumstellar nebulae, the so-called {\it LBV nebulae\/} (LBVN,
e.g. Nota et al.\ 1995).

Only very few LBVs are known in our Galaxy (8 classified and candidate
objects) and a few in other galaxies, for instance in the LMC, SMC, M31, and
M33 (Humphreys \& Davidson 1994). Altogether nearly 40 LBVs and good candidates
are currently known. Not all of them, but many show circumstellar nebulae. In
this paper we analyze the LBVN around the galactic star WRA\,751 and put it
into context with the well known and better studied LBVs $\eta$ Carinae and HR
Carinae.

WRA\,751 ($=$ He 3-591) was first identified as a possible Wolf-Rayet star by
Henize (see Roberts 1962) and then appeared in the Carlson and Henize 
(1979) sample
of southern peculiar emission-line stars, where the authors already recognized
the strong [Fe\,{\sc ii}] lines and classified it as Bep. In addition they
proposed a similarity between WRA\,751 and HR Car (which in the meantime is
known to be a LBV). Using spectroscopic and photometric observations Hu et al.\
(1990) noted for the first time that WRA\,751 is a good LBV candidate. In
addition to the characteristically strong [Fe\,{\sc ii}] lines this star shows
an irregular light variation typical for LBVs. In their analysis the star was
classified as O9.5 with $M_{\rm bol}= -9.6$\mag. They derived $T_{\rm eff} =
30\,000$\,K, $E{\rm (B-V)} = 1.8, M = 50\,{\rm M}_{\sun}$, and a distance of about
5\,kpc.

One year later Hutsem\'ekers \& Van Drom (1991a) found that WRA\,751 was
surrounded by a nebula roughly 22\arcsec\ in diameter. They found the nebula to
be nearly spherically symmetric and to show a diffuse structure. Their
low-dispersion spectra revealed a nebula of low excitation with an electron
temperature of $T_{\rm e,neb} < 15\,000\,$K, an electron density of $n_{\rm
e}\sim 400\,{\rm cm}^{-3}$, and an expansion velocity of $v_{\rm exp} =
26\,$\kms . Assuming a distance of 7\,kpc (derived from radial velocities and
making use of the galactic rotation curve) their radius of the nebula is
0.38\,pc.

Analysis of infrared data (see de Winter 1992) show a NIR excess from which an
estimate for the mass loss of $\dot{M} \sim 10^{-5.5...-6}\,M_{\sun}\,{\rm
yr}^{-1}$ with a velocity of $v_{\rm wind} = 500\,$\kms\ were derived. They
found a cool dusty circumstellar shell with strong emission in the FIR to
surround WRA\,751. In 1992, from interstellar and circumstellar reddening, 
van Genderen measured a distance of $4-5\,$kpc for WRA\,751 as a lower limit. 
Garc{\'\i}a-Lario et al. (1998) re-determined the
characteristic parameters and found $T_{\rm eff} \sim 25\,000\,$K, 
$v_{\rm exp} = 24\,$\kms, and $n_{\rm e} \sim 200\,{\rm cm}^{-3}$.

In summary, the published observations and their interpretations point towards
WRA\,751 being a LBV star with a slowly expanding nebula.

\section{Observation and data reduction}

\subsection{Imaging}

\begin{figure}
{\resizebox{\hsize}{!}{\includegraphics{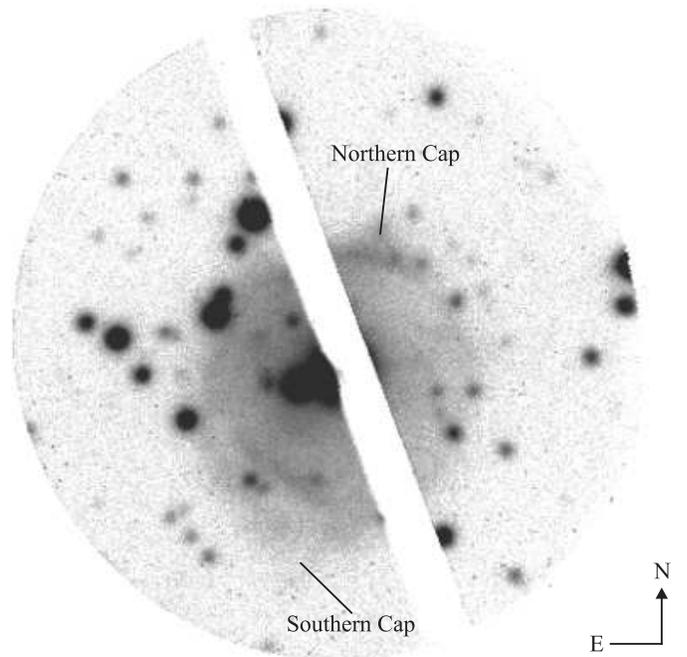}}}
\caption{ESO NTT archive image of the nebula around WRA\,751,
taken with SUSI using an H$_{\alpha}$ filter and a coronographic
mask (see also Nota 1998). The diameter of the field of view
is about 50\arcsec. The Northern and Southern Caps are indicated.
A north-east vector marks the celestial orientation.}
\label{fig:wrantt}
\end{figure}

To compare kinematics and morphology we used images of the nebula
around WRA 751 taken with the {\it Super Seeing Imager\/} (SUSI) on
the {\it New Technology Telescope\/} (NTT) of the {\it European
Southern Observatory\/} (ESO). The images were retrieved from the
ESO NTT archive and reduced in a standard way, using flat-fields
and bias frames from the same observing night. Besides of the
emission-line images in H$_{\alpha}$ and [N\,{\sc ii}] a broad
band red continuum filter image was used for the continuum
subtraction in order to be able to distinguish the pure line
emission from the continuum\footnote{ Due to missing
documentation in the NTT archive and especially the lack of
information in the header of the images a more detailed
specification of the filters can not be given here.}. Fig.
\ref{fig:wrantt} shows a 1050\,s exposure H$_{\alpha}$ image (not
continuum subtracted) with a round field of view, about
50\arcsec\ in diameter. From the image, we determined the seeing
as 0\farcs8. As the image was taken using a coronograph, an
occulting bar obscures the central star and some parts of the
nebula. A non-occulted image of the nebula can be found in
Hutsem\'ekers \& van Drom (1991a).

\subsection{Long-slit echelle spectroscopy\label{sect:echelle}}

To analyze the nebula around WRA\,751 in more detail we preformed
high-resolution long-slit echelle spectroscopy. We used the echelle
spectrograph on the 4\,m telescope at the {\it Cerro Tololo Inter-American
Observatory\/}. Inserting a post-slit H$_\alpha$ filter (6563/75\,\AA) and
replacing the cross-disperser by a flat mirror we selected the H$_{\alpha}$
line and two [N\,{\sc ii}] lines at 6548\,\AA\ and 6583\,\AA. We picked the
79\,l\,mm$^{-1}$ echelle grating and a slit-width of 150\,$\mu$ 
(corresponding to 1\arcsec), which lead to an instrumental 
FWHM at the H$_\alpha$ line of about 10\,km\,s$^{-1}$.

All data were recorded with the long focus red camera and a $2048 \times 2048$
Tek2 CCD.The pixel size was 0.08\,\AA\,pixel$^{-1}$ along the dispersion and
0$\farcs$26\,pixel$^{-1}$ on the spatial axis. Vignetting limited the slit
length to $\sim 4^\prime$ . Seeing was $\sim 2$\arcsec\ during the observations
and the weather was not photometric. Thorium-Argon comparison lamp frames were
taken for wavelength calibration and geometric distortion correction.

\begin{figure}
{\resizebox{\hsize}{!}{\includegraphics{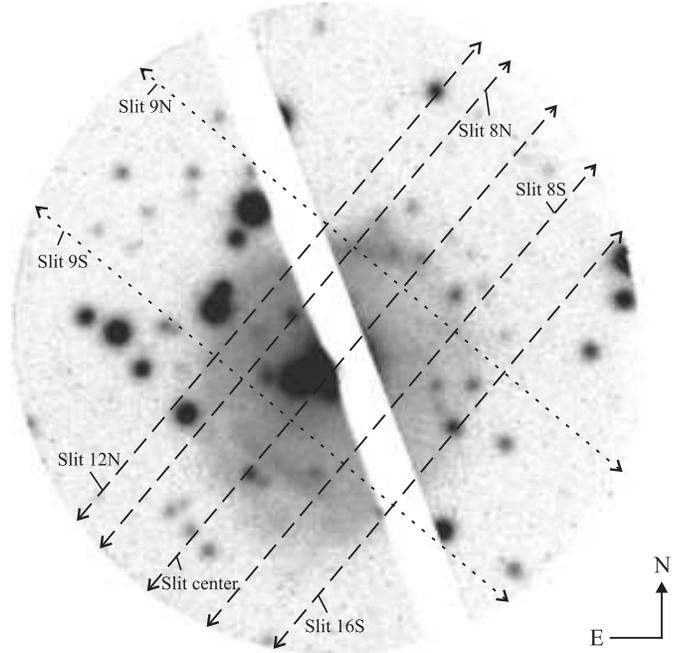}}} \caption{Same image
as in Fig. \ref{fig:wrantt} with the slit positions overlayed: two slits (9N
and 9S) are oriented at ${\rm PA} = 50$\degr, all others at ${\rm PA} =
140$\degr.} \label{fig:wraslits}
\end{figure}

The slit was oriented in two different position angles (PA) perpendicular to
each other (Fig.\ \ref{fig:wraslits}). We observed 5 positions with ${\rm PA} =
140$\degr. One position was centered on the star (in our nomenclature {\it Slit
center\/}), two positions were offset to the north by 8\arcsec\ and 12\arcsec\
from the central star ({\it Slit 8N\/} and {\it Slit 12N\/}) and  two positions
were offset by 8\arcsec\ and 16\arcsec\ south of the central star ({\it Slit
8S\/} and {\it Slit 16S\/}). For ${\rm PA} = 50$\degr\ we observed at two
positions offset from the star by 9\arcsec\ to the north and south,
respectively ({\it Slit 9N\/} and {\it Slit 9S\/}). The left column of Fig.\
\ref{fig:echelle} shows the echellograms of the slits at ${\rm PA} = 140$\degr,
Fig. \ref{fig:echelle50} those at ${\rm PA} = 50$\degr. All echellograms cover
a spectral range of 75\,\AA\, centered on H$_\alpha$ and extend about 2\arcmin\
in the spatial direction, centered on the projected position of the central
star. The top of each echellogram in Fig. \ref{fig:echelle} points to the
north-west, and in Fig. \ref{fig:echelle50} to the south-west. Some telluric
lines are visible and were used to improve the absolute wavelength calibration.

\section{Morphology of the nebula around WRA\,751}

The first image taken of the nebula around WRA\,751 was published
by Hutsem\'ekers \& van Drom (1991a). They discovered a spherical
diffuse nebula about 22\arcsec\ in size. We retrieved and
re-analyzed [N\,{\sc ii}] and H$_{\alpha}$ images retrieved from
the ESO NTT archive in order to compare the morphology and the
kinematics of the nebula. Figure \ref{fig:wrantt} shows that the
nebula around WRA\,751 indeed appears nearly round and almost
spherically symmetric (see also Nota 1998). We measure a
diameter of the main body of the nebula of 22\farcs8, which
corresponds to 0.50\,pc assuming the lower limit distance 
by van Genderen et al.\ (1992; $\sim 4.5\,$kpc).
Appearing rather homogeneous in its surface density and closely
attached to the central star, it resembles very much a Stromgren
sphere. However, this is in contrast to the morphological
appearance of most other LBV nebulae like that of AG Car or HR
Car (e.g.  Hutsem\'ekers \& van Drom 1991b, Nota et al.\ 1995,
Smith et al.\ 1997) which show a more intense detached ring
structure, or as in the case of HR Car at least parts of a bipolar shell
(Weis et al. 1997, Nota et al. 1997). 
This is mainly due to an increased matter density along
the line of sight when looking through the edge of a shell.

A closer inspection of the images allows us to locate spatial variations in the
surface density of the nebula around WRA\,751: a brighter circular half-shell
shows up in the eastern part, while---in comparison---in the western part the
brightness decreases by about a factor of two. Even more remarkable are two
extensions of the spherical shell, one nearly exactly to the north at a
position angle of 340\degr\ which we will call the {\it Northern Cap\/} (Fig.\
\ref{fig:wrantt}) and the other to the south ({\it Southern Cap\/}) at ${\rm
PA} = 165$\degr. This means that the Caps are almost along an axis through the
star. Morphologically both appear roughly triangular in shape. The surface
brightness of the Northern Cap is somewhat larger, that of the Southern Cap
somewhat smaller than the low surface part (the west side) of the central
spherical nebula. The Northern Cap extends by 2\farcs6 beyond the main body of
the nebula, the Southern Cap by 4\farcs7.

\section{Kinematic indication of a bipolar structure}

\begin{figure*}
\begin{center}
{\resizebox{14.3cm}{!}{\includegraphics{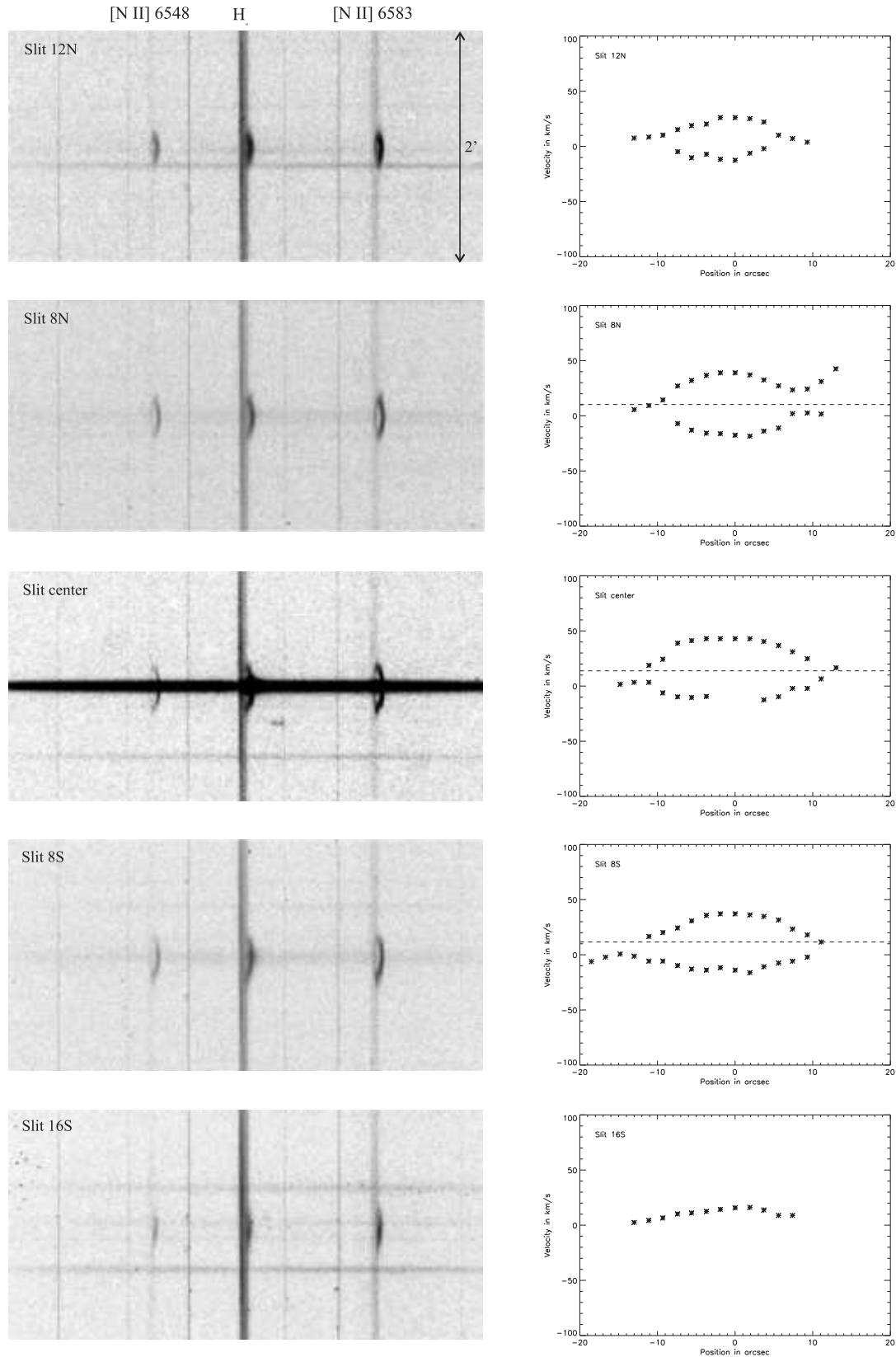}}}
\end{center}
\caption{Echellograms (left column) and corresponding position-velocity
diagrams (right column) for the slits with ${\rm PA} = 140$\degr. The positions
are centered on the projected location of the star, the velocities are with
respect to the LSR. Dashed lines in the $pv$-diagrams mark the center of
expansion velocities.} \label{fig:echelle}
\end{figure*}

\begin{figure*}
\begin{center}
{\resizebox{15cm}{!}{\includegraphics{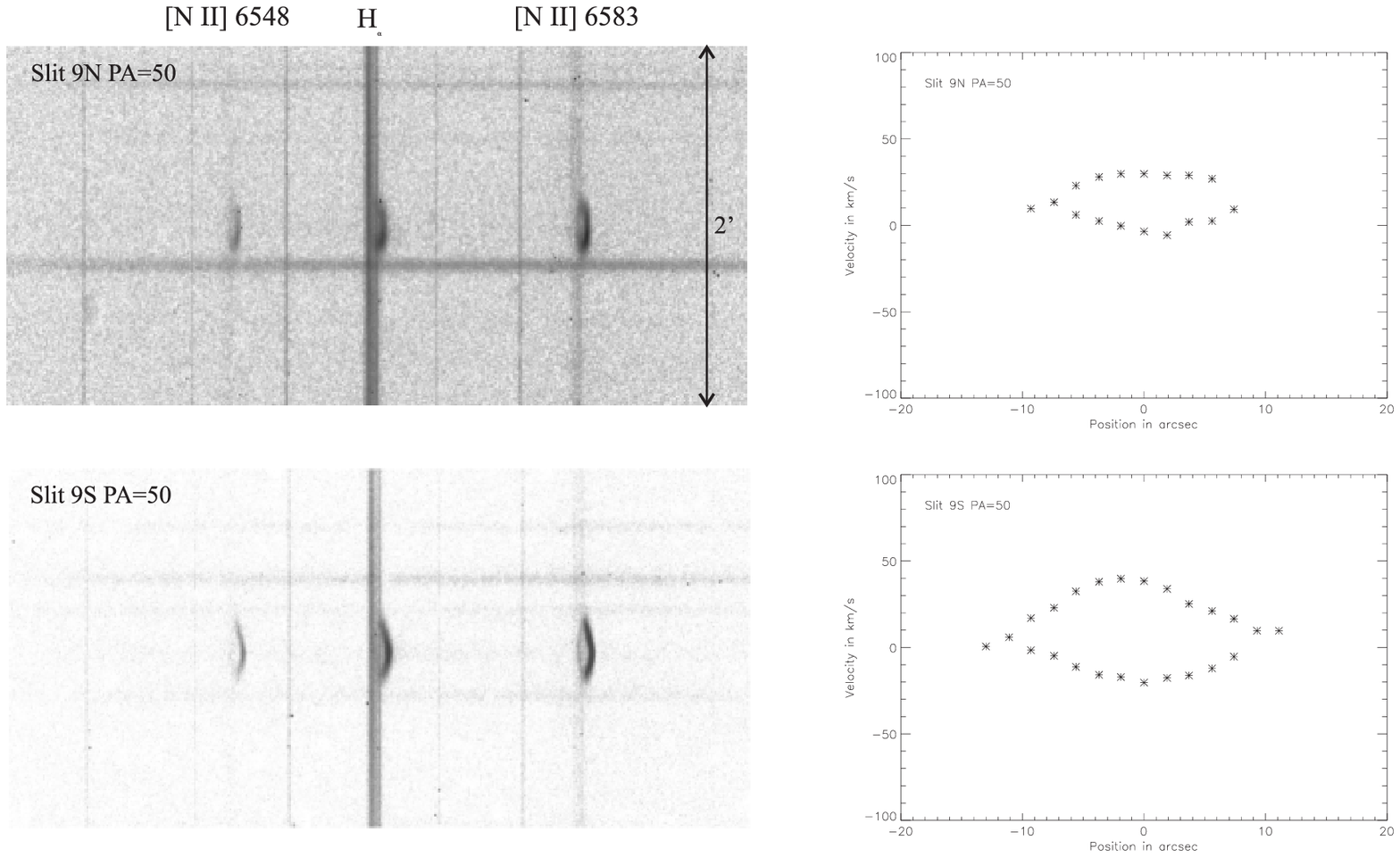}}}
\end{center}
\caption{Echellograms (left column) and corresponding position-velocity
diagrams (right column) for the slits with ${\rm PA} = 50$\degr. The positions
are centered on the projected location of the star, the velocities are with
respect to the LSR.} \label{fig:echelle50}
\end{figure*}

Using high-resolution long-slit echelle spectra an analysis of the
kinematics of the LBVN around WRA\,751 was performed. With the 7
slit positions our mapping covers the entire nebula. The right
columns of both, Figs.\ \ref{fig:echelle} and \ref{fig:echelle50},
show the position-velocity diagrams ($pv$-diagrams) corresponding
to the echellograms in the same line. For the $pv$-diagrams we
used the stronger [N{\sc ii}] line at 6583\AA\ which is less
contaminated by background emission. Additionally all measurements
were compared to those of H$_{\alpha}$ to ensure that no kinematic
difference exists between the two lines. All velocities were
corrected to the {\it Local Standard of Rest\/} (LSR) system.  The
0\arcsec\ position in all $pv$-diagrams corresponds to the
projected position of the central star (WRA\,751) onto the slit,
from there positive offsets are to the north-west for the ${\rm
PA} = 140$\degr\ slits (or south-west for ${\rm PA} = 50$\degr)
and negative to the south-east (or north-east, respectively). All
spectra (except Slit 16S) clearly resolve the Doppler ellipse,
and indicate a predominantly spherical expansion of the nebula. The
largest radial expansion velocity $v_{\rm exp}$ was found in the
central position (0\arcsec) of Slit 9S (${\rm PA} = 50$\degr) with
a value of 29.4\,\kms $\pm$ 2\,\kms, the second largest in Slit 8N
at 28.3\,\kms $\pm$ 2\,\kms. A similar expansion velocity was
derived at the same position in the central slit at 26.3\,\kms
$\pm$ 2\,\kms. Since this slit crosses the star in the center, the
velocity measurements are less certain at the 0\arcsec\ position.
There $v_{\rm exp}$ might be even higher as can be estimated from
interpolating for the missing data points in the $pv$-diagram of
Fig.\ \ref{fig:echelle}. If a nebula is spherically expanding, the
largest expansion velocity should occur at the position
projected onto the star. Even though this is not the case here,
the difference of the expansion velocity between the central slit
and the Slit 8N is not significant and within the errors.

However, a comparison with Slit 9S (${\rm PA} = 50$\degr) shows 
an asymmetric shape of the expansion structure indicating that 
the nebula is most likely not perfectly round but
perhaps a tilted ellipsoid with the maximum expansion off-centered
to the east of WRA\,751. There the largest expansion
velocities are found. Another possibility is that the nebula 
is spherical with a bump at
this point. This is best visible in the asymmetric shapes of
the expansion ellipses in Slits 8N (Fig.\ \ref{fig:echelle}) and 
Slit 9S (${\rm PA} = 50$\degr; Fig.\ \ref{fig:echelle50}). While the
largest negative velocity is found at the 0\arcsec\ position, the
largest positive velocity is located at $-2$\arcsec, i.e., the
Doppler ellipse is asymmetric with respect to the projected
position of the star. All derived global expansion velocities are
in good agreement with expansion velocities found in the
literature (Hutsem\'ekers \& van Drom 1991a, Garc{\'\i}a-Lario et
al.\ 1998).

In almost all spectra the redshifted line is more intense and
broader than the blueshifted side. Since it is the redshifted
component which is stronger, this brightness difference cannot be
accounted by an absorption of the blueshifted component. It is
either an intrinsic brightness difference or results from a
thicker shell at the backside which leads to a higher emission
measure. The FWHM of the redshifted wing of the expansion ellipse
is resolved at 21\,\kms\ (instrumental FWHM $\sim10\,$\kms) and
supports the interpretation as a thicker shell. The widths itself
is indicative of a stratification of the radial velocities along a
line of sight, presumably with the inner parts of the nebula
moving slower. A wider redshifted component can be explained as
due to a thicker shell there.

\begin{figure}
{\resizebox{\hsize}{!}{\includegraphics{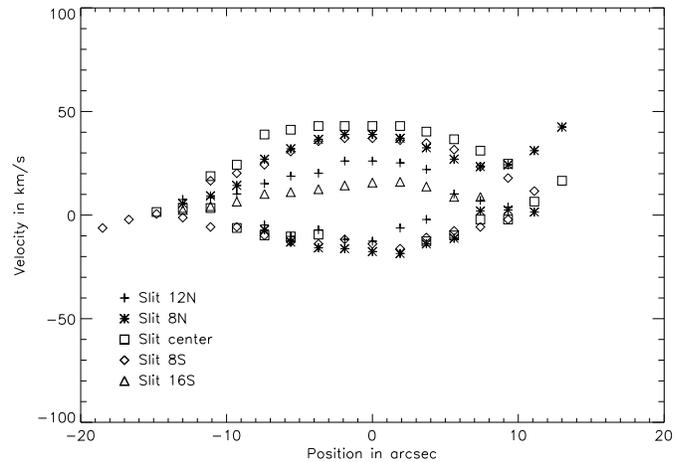}}} \caption{Overlay of all
position velocity diagrams at ${\rm PA} = 140$\degr.} \label{fig:pvall}
\end{figure}

Fig.\ \ref{fig:pvall} shows a composite $pv$-diagram of all slits with ${\rm PA} =
140$\degr. It shows the trend of the expansion velocities across the nebula:
\begin{itemize}
\item The split of the Doppler ellipse decreases as we move away from the
geometric center of the bubble. The decrease, however, is not in agreement with
a purely spherical expansion. While the redshifted part of the expansion
ellipse decreases more like that of a spherical expansion, the blueshifted
component stays nearly at a constant velocity for a given position along the
slit.
\item For a spherical expansion not only the width of the expansion ellipse in
velocity space shrinks, but it also becomes narrower in spatial extent. In the
WRA\,751 nebula, however, we find that the convergence points at positive
positions migrate to smaller values as one proceeds to slits farther away from
the star while they are at approximately the same location for negative values.
\end{itemize}
Both points are in agreement with an asymmetry within the nebula and indicate
deviations from a spherical expansion.

Besides the---asymmetric---expansion of the shell, the spectra show
additional kinematic components. At slit position 8N the redshifted side
converges with the blueshifted side in the north-west, at the position of about
10\arcsec\ as expected at this position for an expanding bubble. At the same
point, however, a component which is more redshifted as the nebula appears, and
extends to a position of 13\arcsec. The redshift of this component increases to
32\,\kms above the central expansion velocity as one moves away from the
nebula. In Fig.\ \ref{fig:echelle} a dashed line in the $pv$-diagrams for Slits
8N, center, and 8S indicates the rest velocity of the kinematic center of the
expansion at 10\,\kms, 14\,\kms, and 12\,\kms, respectively.

A similar kinematic behavior can be found in Slits central and 8S, but this
time at the south-eastern rim of the nebula. At about the position
$-12$\arcsec\ in both spectra an additional blueshifted component appears with
maximum velocities of 12 and 18\,\kms\ below the respective central expansion
velocities. Again the velocities of these components reach their maximum at the
location most distant from the star.

The position of the two kinematic extensions is very symmetric with respect to
the central star and thus indicative of a bipolar kinematic structure. At Slit
8N material at the rim of the inner (nearly) spherical nebula moves away from
us while in Slit 8S material at the corresponding positions, namely the edge
of the nebula, approaches us. With respect to the center of expansion of the
central spherically expanding nebula these kinematic components are bipolar.

The spatial distribution of the velocity field corresponds well to the
morphology discussed above: The main part of the expansion (between the two
convergence points of the expansion ellipse) measures 23\arcsec\ in diameter
and thus agrees with the diameter of the main body of the nebula as determined
from the image to be 22\farcs8. Moreover, it is noteworthy that the sizes of
the kinematic extensions as measured from the spectra and of the Northern and
Southern Caps as determined from the images agree equally well.

To ensure that the Caps are part of the nebula around WRA\,751 and not just a
background object onto which the nebula is projected, we have determined the
line ratio [N\,{\sc ii}]\,6583\AA/H$_\alpha$ after subtracting contributions of
the telluric H$_\alpha$ line. Within the accuracy of our measurements, the
entire nebula, including both Caps, shows a constant value of [N\,{\sc
ii}]\,6583\AA/H$_\alpha = 1.2 \pm 0.1$, in contrast with the value for the
background material of 0.3. The background line ratio is compatible with the
value for a galactic H\,{\sc ii} region (Shaver et al.\ 1983). This difference
ensures that we can disentangle source and background contributions.

\section{Discussion and conclusions}

A comparison of detailed kinematic data and the morphology of the LBVN around
the galactic star WRA\,751 indicates that the expansion is not as perfectly
spherically symmetric as previously thought. The redshifted central part of the
nebula expands nearly spherically while the blueshifted part appears stalled
and moves at an almost constant radial velocity of about $-15$\,\kms (Fig.\
\ref{fig:pvall}). A natural reason may be that the front side of the nebula
expands into a denser medium and that thus the expansion is decelerated. The
maximum expansion velocity was found off-centered from the central star adding a
further piece of evidence that the nebula is neither exactly spherical nor
expanding accordingly. The off-centered maximum velocity is either an evidence
for an elongated (elliptical) shape and tilt or for a bump on the back side of
the nebula.

Two kinematic extensions, one redshifted, one blueshifted appear
attached to the ridge of the nebula. A comparison with the
morphology of the nebula around WRA\,751 allows to identify the
redshifted extension (Slit 8N) with the Northern Cap and the
blueshifted one (Slits center and 8S) with the Southern Cap (see
Cap and Slit positions in Figs. \ref{fig:wrantt}\ and
\ref{fig:wraslits}). The Northern and the Southern Cap therefore
indicate morphologically and kinematically bipolar
components in the nebula around WRA\,751. While one may speculate
about a jet like structure or---similar to planetary nebulae (for
a discussion of similarities between LBVs and planetary nebulae,
see, e.g., Frank 1998)---wider funnels, or an altogether
elliptical form of the nebula creating the bipolar appearance, due
to the limited spatial resolutions of both the available kinematic
data and the available images, higher resolution observations will
be necessary to further constrain the structure of the nebula
around WRA\,751.

Nonetheless, from the data presented one can draw several conclusions:
\begin{itemize}
\item The LBVN around WRA\,751 consists of an expanding shell of 23\arcsec\ in
diameter, with a thicker shell at the back side. The radial
velocity distribution deviates from that of a simple spherical
expansion pattern. We find bipolar contributions to the
morphology as well as radial velocity distribution in
approximately north-south direction (Northern and Southern Caps).
\item The maximum expansion velocities are off-centered which most 
likely can be accounted for by a bump in the back side east of the 
central star.
\item The [N\,{\sc ii}]\,6583\AA/H$_\alpha$ ratio is considerably 
larger than that
of the background material. This is typical for CNO-processed material.
\end{itemize}
These three results are fully consistent with WRA\,751 being a
true member of the LBV class of stars. In many of the well
investigated ones bipolar components to the morphological as well
as kinematic structures of the nebula are found. $\eta$ Car
(Duschl et al.\ 1995, Morse et al.\ 1998) and HR Car (Weis et al.\
1997, Nota et al.\ 1997) are the most prominent examples. As stars
in late phases of their evolution, LBVs quite naturally show
CNO-processed material in their envelopes and ejecta
(Garc{\'\i}a-Segura et al.\ 1996, Smith et al.\ 1978). A
large [N\,{\sc ii}]\,6583\AA/H$_\alpha$ ratio in a nebula is often
used as one of the indicators for a star being a LBV. One finds,
for instance, values of [N\,{\sc ii}]\,6583\AA/H$_\alpha = 3\
\dots\ 7$ for $\eta$ Car (Davidson et al.\ 1982, Meaburn et al.\
1987, 1996, and Weis et al.\ 1999), $= 0.4\ \dots\ 0.9$ for HR Car
(Hutsem\'ekers \& van Drom 1991b, and Weis et al.\ 1997), and $=
0.7$ for AG Car (Thackeray 1977 and Smith et al.\ 1997).

Combining our morphological, kinematic and spectroscopic
results, we find mounting evidence that WRA\,751 is a LBV,
indeed. Moreover, there are bipolar components in its nebula,
notably the two Caps. While the bipolarity in this object is less
pronounced and less obvious than in other LBVs, it still strengthens
our suspicion that bipolarity---albeit at different levels---is a
genuine property of LBV nebulae. This makes it even more likely
that modeling LBVNs as simple windblown spheres is an unjustified
oversimplification.

\begin{acknowledgements}
The author is very grateful to Wolfgang J.\ Duschl and Dominik J.\ Bomans for
many discussions on the subject of this paper, and for carefully reading
and improving the manuscript. Sincere thanks go to You-Hua Chu for her
support. Part of the work was carried out as a visiting graduate student
at the Department of Astronomy of the University of Illinois. Its
hospitality is gratefully acknowledged. The author thanks the referee, 
Damien Hutsem\'ekers for helpful comments on the paper. 

\end{acknowledgements}

\end{document}